\documentclass[floatfix,twocolumn,english,aps,prb,superscriptaddress]{revtex4-1}
\usepackage{grffile}
\RequirePackage{fix-cm} % arbitrary font scaling
\usepackage{siunitx}
\usepackage{gensymb}
\usepackage[latin9]{inputenc}
\usepackage{color}
\usepackage{babel}
\usepackage{amssymb}
\usepackage{amsfonts}
\usepackage{graphicx}
\usepackage{epstopdf}
\usepackage{multirow}
\usepackage{cancel}
\PassOptionsToPackage{normalem}{ulem}
\usepackage{ulem}
\usepackage[version=3]{mhchem}
\usepackage[unicode=true, bookmarks=false, breaklinks=false,pdfborder={0 0 1},colorlinks=false] {hyperref}
\usepackage[T1]{fontenc}
\begin{document}

\title{Mapping the three-dimensional fermiology of the triangular lattice magnet \ce{EuAg4Sb2}} 

\author{J. Green\textdagger}
\affiliation{Department of Physics and Astronomy and California NanoSystems Institute, University of California, Los Angeles, CA 90095, USA}
\thanks{These authors contributed equally to this work.}

\author{Harry W. T. Morgan\textdagger}
\affiliation{Department of Chemistry and Biochemistry, University of California, Los Angeles, CA 90095, USA}
\thanks{These authors contributed equally to this work.}

\author{Morgaine Mandigo-Stoba}
\affiliation {Department of Physics and Astronomy, University of California, Los Angeles, CA 90095, USA}

\author{William T. Laderer}
\affiliation{Department of Chemistry and Biochemistry, University of California, Los Angeles, CA 90095, USA}

\author{Kuan-Yu Wey}
\affiliation{Department of Physics and Astronomy, University of California, Los Angeles, CA 90095, USA}

\author{Asari G. Prado}
\affiliation{Department of Physics and Astronomy, University of California, Los Angeles, CA 90095, USA}

\author{Chris Jozwiak}
\affiliation{Advanced Light Source, E.O. Lawrence Berkeley National Laboratory, Berkeley, CA, USA}

\author{Aaron Bostwick}
\affiliation{Advanced Light Source, E.O. Lawrence Berkeley National Laboratory, Berkeley, CA, USA}

\author{Eli Rotenberg}
\affiliation{Advanced Light Source, E.O. Lawrence Berkeley National Laboratory, Berkeley, CA, USA}

\author{Christopher Guti\'errez}
\affiliation {Department of Physics and Astronomy, University of California, Los Angeles, CA 90095, USA}

\author{Anastassia N. Alexandrova}
\affiliation{Department of Chemistry and Biochemistry, University of California, Los Angeles, CA 90095, USA}

\author{Ni Ni}
\email{Corresponding author: nini@physics.ucla.edu}
\affiliation {Department of Physics and Astronomy and California NanoSystems Institute, University of California, Los Angeles, CA 90095, USA}

\begin{abstract}
In this paper, we report the temperature-field phase diagram as well as present a comprehensive study of the electronic structure and three-dimensional fermiology of the triangular-lattice magnet \ce{EuAg4Sb2}, utilizing quantum oscillation measurements, angle-resolved photoemission spectroscopy, and first-principles calculations. The complex magnetic phase diagram of \ce{EuAg4Sb2} highlights many transitions through nontrivial AFM states. Shubnikov-de Haas and de Haas-van Alphen oscillations were observed in the polarized ferromagnetic state of \ce{EuAg4Sb2}, revealing three pairs of distinct spin-split frequency branches with small effective masses.
A comparison of the angle-dependent oscillation data with first-principles calculations in the ferromagnetic state and angle-resolved photoemission spectra shows good agreement, identifying tubular hole pockets and hourglass-shaped hole pockets at the Brillouin zone center, as well as diamond-shaped electron pockets at the zone boundary. 
As the temperature increases, the frequency branches of the tiny hourglass pockets evolve into a more cylindrical shape, while the larger pockets remain unchanged. This highlights that variations in exchange splitting, driven by changes in the magnetic moment, primarily impact the small Fermi pockets without significantly altering the overall band structure. This is consistent with first-principles calculations, which show minimal changes near the Fermi level across ferromagnetic and simple antiferromagnetic states or under varying on-site Coulomb repulsion.

\end{abstract}
\date{\today}
\maketitle

\section{Introduction}

Magnetic topological materials have attracted considerable research interest due to the intriguing interplay of magnetism, topology, and charge carriers. One strategy for discovering new magnetic topological materials is doping a known nonmagnetic topological material with magnetic ions, or even replacing the nonmagnetic elements with magnetic ones. Although, in many cases, this approach fails to yield new magnetic topological phases due to the strong electron correlations and the complexity introduced by magnetism, it has led to the discovery of ferromagnetic (FM) topological insulator \ce{Cr_{0.15}(Bi_{0.1}Sb_{0.9})_{1.85}Te3} and the quantum anomalous Hall effect \cite{QAH}. 

Our recent study of quantum oscillation and the first-principles calculations of \ce{SrAg4Sb2} shows that when spin-orbit Coupling (SOC) is considered, a band inversion appears at the Brillouin zone boundary T point, making \ce{SrAg4Sb2} a topological crystalline insulator candidate \cite{morgan-SrAg4Sb2-bonding, Sr142}. 
\ce{SrAg4Sb2} crystallizes in a \ce{CaCu4P2}-type structure with space group $R\bar{3}m$ and belongs to a big family of \ce{EuT4X2} (T = Cu and Ag; X = P, As and Sb) \cite{EuAg4As2-Sb2, HyperfineSplitting, EuCu4As2, EuCu4As2-2, EuCu4P2}. Its magnetic analog, \ce{EuAg4Sb2} where Sr is replaced by magnetic Eu$^{2+}$, shows two weakly first-order magnetic phase transitions at $T_1 =$ 10.6 K and $T_2 =$ 7.5 K. Recently, large magnetoresistance was observed in \ce{EuAg4Sb2}, suggesting strong interplay between magnetism and charge transport \cite{EuAg4Sb2}. Large magnetoresistance was also observed in its sister compounds \ce{EuAg4As2} and \ce{EuCu4As2} \cite{Shen2018, Zhu2020, EuAg4As2-Pressure, LargeMR-EuCu4As2}. Neutron diffraction studies on \ce{EuAg4As2} indicate incommensurate noncollinear antiferromagnetism with a propagation vector of (0, 0.1, 0.12) where the magnetic structure was revealed to be helical along the $c$ axis and cycloidal along the $b$ axis \cite{Shen2018}. Therefore, it will be of particular interest to elucidate the fermiology and topological nature of the magnetic compounds in this family.

In this paper, we report the temperature-field phase diagram of \ce{EuAg4Sb2} and observation of quantum oscillations, as well as our thorough investigation of the electronic band structure and the three-dimensional (3D) fermiology in \ce{EuAg4Sb2}. Through the analysis of angle-dependent and temperature-dependent Shubnikov-de Haas (SdH) and de Haas-van Alphen (dHvA) quantum oscillations (QO), and comparison of these findings with Density Functional Theory (DFT) calculations and Angle Resolved Photoemission Spectroscopy (ARPES) measurements, we reveal the spin-split Fermi surfaces with light effective masses and identify hourglass-shaped hole pockets and tube-like hole pockets at the center of the Brillouin zone, along with diamond-shaped electron pockets located at the zone boundary. Band structure calculations show no significant variations near the Fermi level between ferromagnetic (FM) and simple antiferromagnetic (AFM) configurations or with changes in on-site Coulomb repulsion, classifying \ce{EuAg4Sb2} as a weakly-correlated AFM semimetal. 

\begin{figure}

    \centering   \includegraphics[width=\columnwidth]{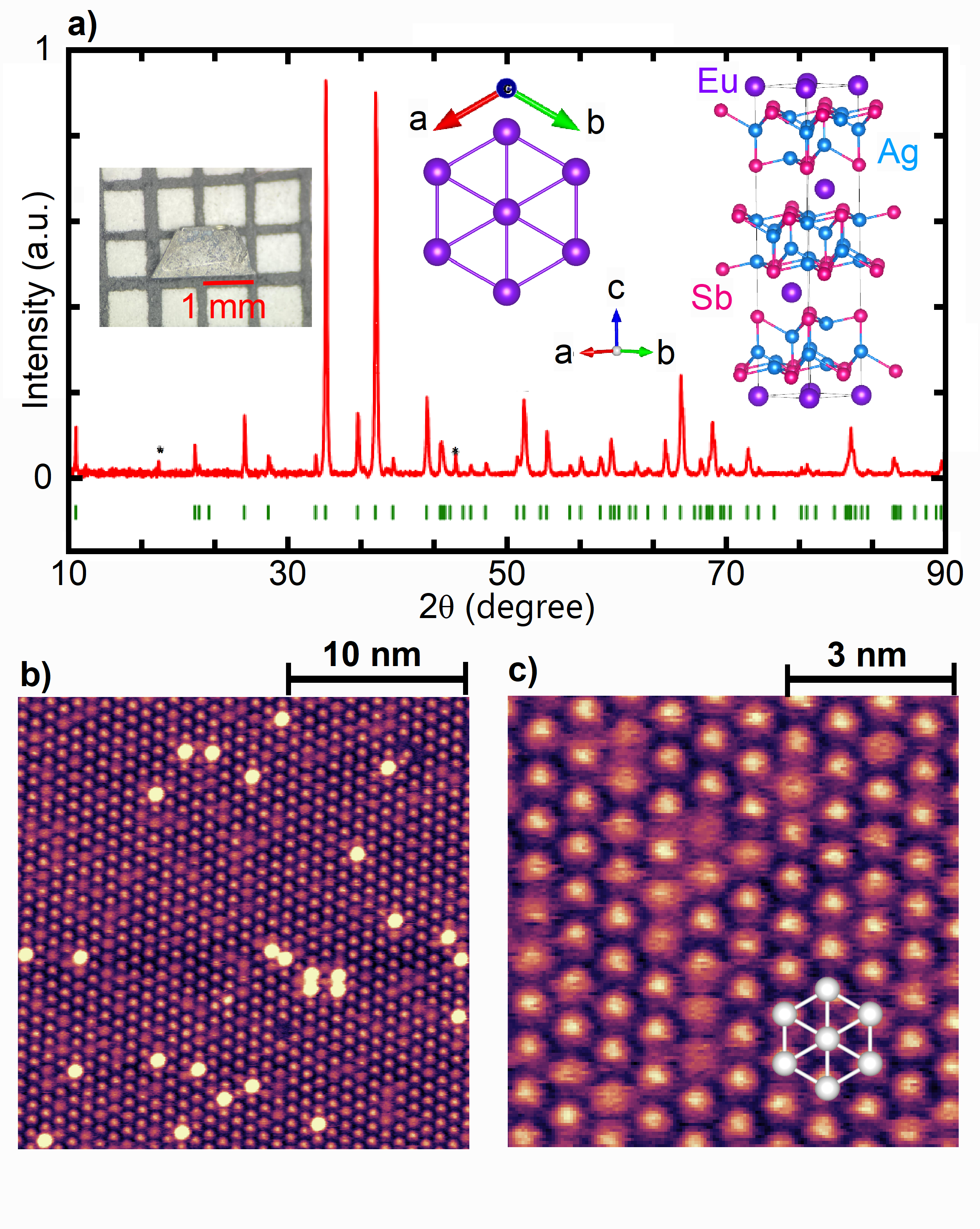}
    \caption{(a) Powder x-ray diffraction pattern of \ce{EuAg4Sb2}. Asterisks indicate small impurity peaks of unreacted Sb and Eu. Green tick marks indicate the calculated Bragg peak positions. The left inset shows a picture of a single crystal against a 1 mm - sized grid and the right insets are a rendering of the atomic configuration within the crystal. (b) 25$\times$25 nm constant-current STM image taken at 77 K of \ce{EuAg4Sb2} ($I_T$ = 8 pA, $V_B$ = 500 mV). (c)  8$\times$8 nm zoom image of the surface. Overlaid on the atoms is a rendering of the triangular $(\sqrt{3}\times\sqrt{3})R30^\circ$ surface lattice reconstruction which describes a superlattice where the unit cell is $\sqrt{3}$ times larger than the original $ab$ lattice plane in both dimensions and is rotated by $30^\circ$ relative to the underlying plane.}
    \label{xray}
    
\end{figure}

\section{Experimental methods}
\ce{EuAg4Sb2} single crystals were first synthesized using the self-flux method. Eu, Ag, and Sb pieces were mixed according to the molar ratio 1 : 10 : 5 and placed inside an alumina crucible, which was then placed in a quartz tube. The quartz tube was evacuated and then heated to 1100$\celsius$ and held at that temperature for 5 hours to ensure homogenization. Subsequently, it was cooled to 600$\celsius$ over 160 hours. The flux was then separated from the crystals using a centrifuge. However, we found that when cooled from this high temperature, there exists a second stable phase, \ce{EuAgSb}. Much like the non-magnetic counterpart \ce{SrAg4Sb2}, \ce{EuAgSb} formed in between two layers of \ce{EuAg4Sb2}, complicating the synthesis of single crystals \cite{Sr142}. To avoid the introduction of \ce{EuAgSb}, we heated the growth ampule to 1100$\celsius$ to guarantee homogenization, then rapidly cooled to 800$\celsius$ at 200$\celsius$/hr. The ampule was held at 800$\celsius$ for 3 hours before being slowly cooled to 580$\celsius$ at a rate of 2$\celsius$/hr. Then single crystals were separated from the liquid flux by a centrifuge. Initiating the cool-down process at 800$\celsius$ resulted in a high yield of free-standing \ce{EuAg4Sb2} single crystals, characterized by a long bar-shaped morphology and hexagonal plate-like samples, as illustrated in the inset of Fig. \ref{xray}.

Magnetic susceptibility measurements were performed in a QD VSM Magnetic Property measurement system (MPMS3) in a maximum field of 7 T. Magnetotransport and magnetic torque measurements were performed in a Quantum Design Dynacool Physical Property Measurement System up to 9 T. Torque measurements were made by mounting a small piece of single crystal on the tip of a piezoresistive self-sensing silicon cantilever. The magnetic torque was then inferred from the magnetoresistance of the cantilever measured by a Wheatstone bridge, as the resistance of the cantilever is very sensitive to the deformation caused by torque. Hall ($\rho_{yx}$) resistivity measurements were performed using the six-probe technique. To eliminate unwanted contributions from mixed transport channels, data were collected while sweeping the magnetic field from -9 T to 9 T. The data were then antisymmetrized to obtain $\rho_{yx}(B)$.

ARPES experiments were performed at the $\mu$-ARPES endstation of the MAESTRO beamline at the Advanced Light Source (ALS) at Lawrence Berkeley National Lab (LBNL). Samples were cleaved in-situ under ultra-high vacuum (UHV) conditions ($<{2\times 10^{-11}}$ Torr) at temperature $T \approx 10$ K. 
Measurements were conducted at a base temperature of 10 K using linearly polarized light with photon energies of $h\nu = 100$ eV and 120 eV. The beam size was 40 $\mu$m $\times$ 63.4 $\mu$m, and data were collected with a Scienta R4000 analyzer. Angular and energy resolution were set to approximately 0.1 - 0.3 $\degree$ and 62.5 meV, respectively. ARPES data were processed using open-source Python library PyARPES. \cite{pyARPES}. Scanning Tunneling Microscopy (STM) experiments were performed at temperature $T= 77.4$ K and UHV conditions (pressure $\sim 10^{-10}$ Torr) using a PtIr tip operated in constant current mode. 

The electronic structure of \ce{EuAg4Sb2} was studied via DFT calculations using the PBE functional\cite{hohenberg,kohn,pbefunctional} and the projector augmented wave (PAW) pseudopotential method as implemented in the Vienna Ab initio Simulation Package (VASP), version 5.4.4 \cite{blo}.
The DFT+U method was used to account for strong electronic correlation within the $f$ orbitals of Eu. 
An approximate SOC correction implemented in VASP was used to compute the electronic properties in the first Brillouin zone \cite{vasp soc}.
Fermi surfaces were computed using a $k$-mesh spacing of 0.008 (31-31-31).
Fermi surface data for visualization were generated from DFT calculations using Vaspkit and visualized with Fermisurfer \cite{vaspkit,fermisurfer}.
Fermi surface data for calculation of dHvA frequencies were generated from VASP using c2x.\cite{c2x}
dHvA frequencies were computed using SKEAF\cite{skeaf}.
Band structures were plotted with pyprocar version 5.6.6 \cite{pyprocar}.

\begin{figure}

    \centering
    \includegraphics[width=\columnwidth]{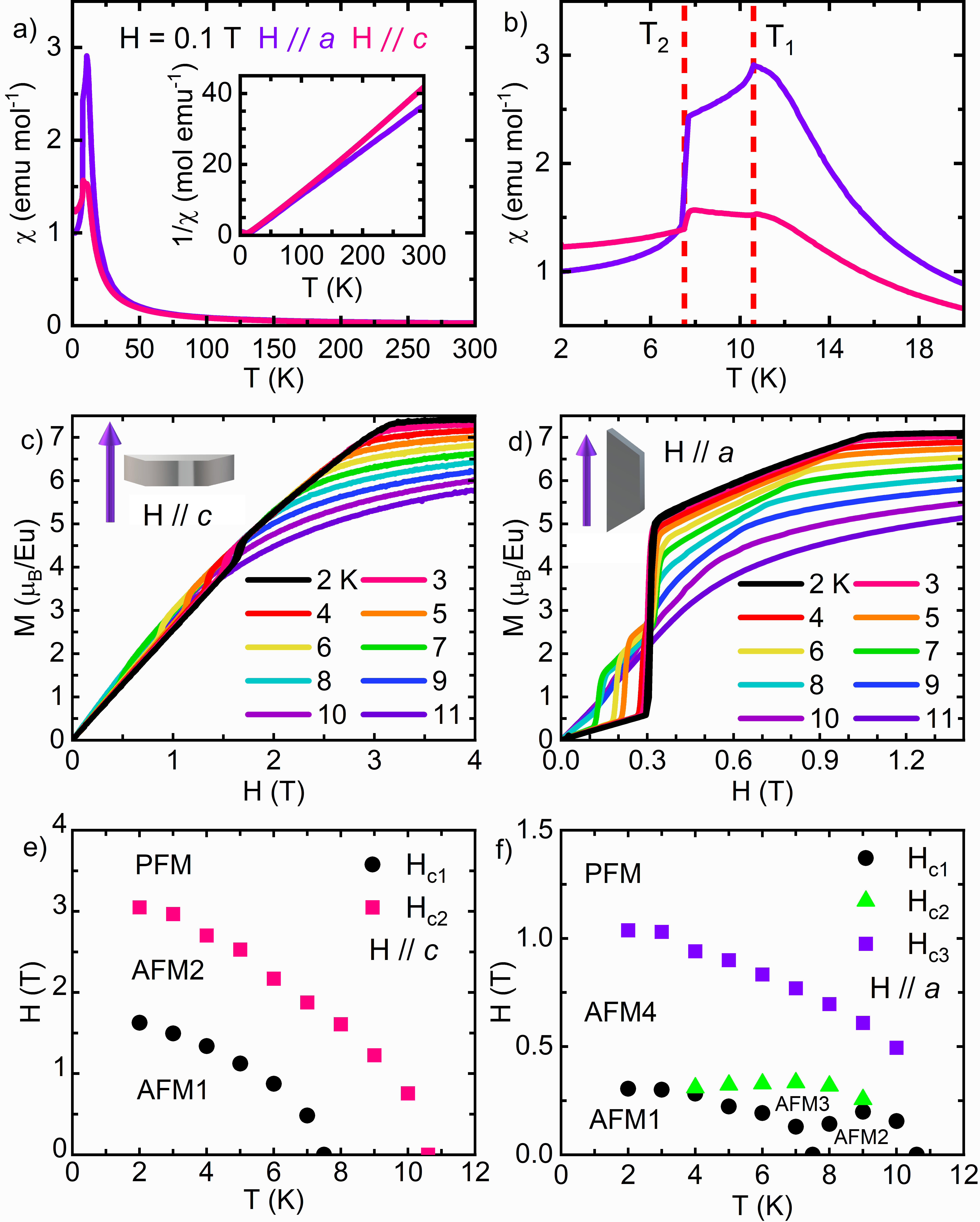}
    \caption{The magnetic properties of \ce{EuAg4Sb2}. (a) Temperature-dependent magnetic susceptibility, $\chi(T)$, measured under an applied field of 0.1 T along $H\parallel c$ (pink) and $H\parallel a$ (purple) from 300 K to 2 K. The inset shows the inverse susceptibility, 1/$\chi(T)$. (b) A zoomed-in view of $\chi(T)$ for temperatures below 20 K. (c) and (d) Isothermal magnetization, $M(H)$, measured at various temperatures for $H\parallel c$ and $H\parallel a$, respectively, revealing multiple metamagnetic transitions. (e) and (f) Temperature-field phase diagrams for $H\parallel c$ and $H\parallel a$, respectively. The critical fields were determined from the derivatives of $M$ with respect to $H$ (d$M$/d$H$). These phase diagrams align well with those reported in Ref. 9 despite slight discrepancy observed in the phase diagram for $H\parallel c$ which may be due to a minor misalignment of the magnetic field. }
    \label{property}
    
\end{figure}

\begin{figure*}

    \centering
    \includegraphics[width=\textwidth]{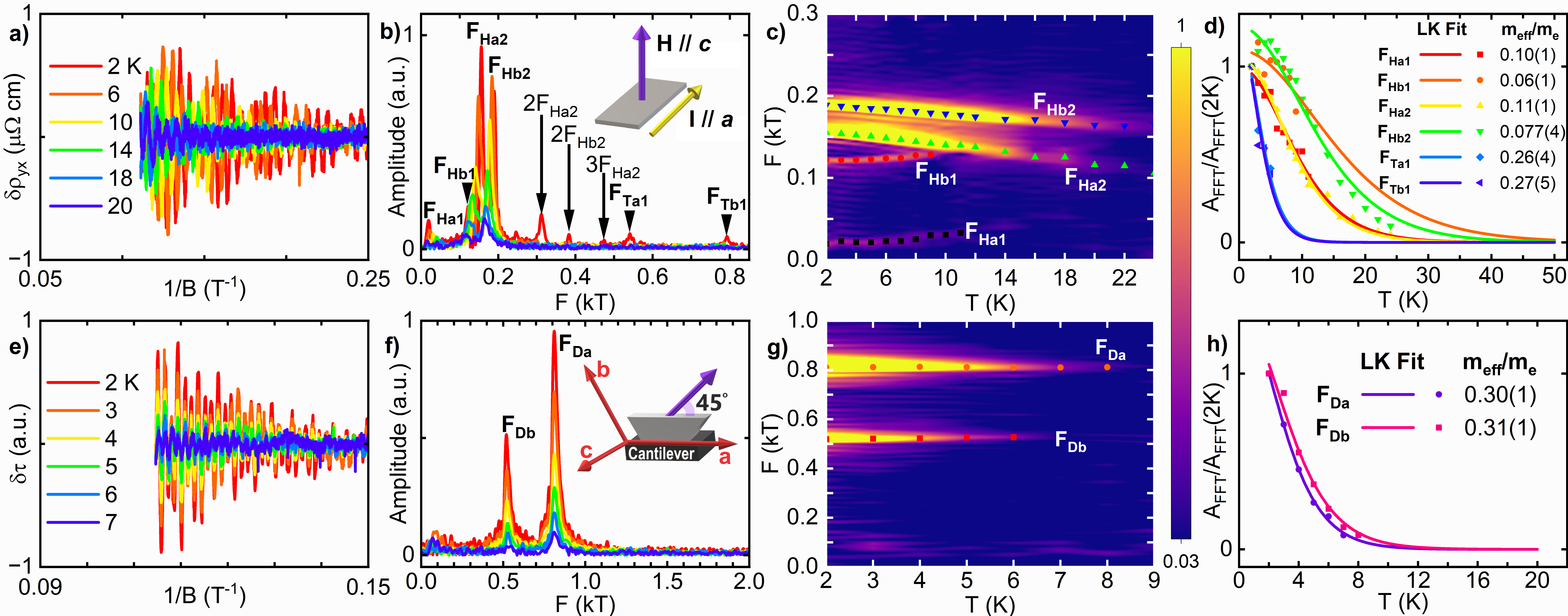}
    \caption{(a) $\delta \rho_{yx}$ at temperatures 2 - 20 K as a function of $1/B$. (b) The FFT curves associated with (a). The top right inset illustrates the measurement configuration. (c) A contour plot of the FFT curves as a function of temperature. (d) The normalized temperature-dependent FFT amplitude. The LK fits used to extract the effective masses are shown as the solid lines. (e) $\delta \tau$ at temperatures 2 - 7 K as a function of $1/B$ when the magnetic field is in the $ab$ plane. (f) The FFT curves associated with (e). The top right inset illustrates the measurement configuration. (g) A contour plot of the FFT curves as a function of temperature. (h) The normalized temperature-dependent FFT amplitude. The LK fits used to extract the effective masses are represented as solid lines.}
    \label{QOTemp}
    
\end{figure*}

\section{Experimental Results}  
\subsection{Characterization and physical properties of \ce{EuAg4Sb2}}

\ce{EuAg4Sb2} crystallizes in the \ce{CaCu4P2}-type centrosymmetric trigonal space group $R\bar{3}m$ (No. 166) with lattice parameters $a$ = $b$ = 4.7427(5) \AA, $c$ = 24.6979(3) \AA, $\alpha$ = $\beta$ = 90$\degree$ and $\gamma$ = 120$\degree$. The crystal structure at room temperature is shown in the inset of Fig. \ref{xray}(a). In this compound, layers of \ce{Ag4Sb2} are separated by Eu atoms, where the Eu atoms are arranged in a triangular lattice. Each Eu atom has six nearest Eu neighbors, creating the potential for geometric frustration. Figure \ref{xray} (a) shows the x-ray diffraction pattern of the powder obtained from several single crystals. Green tick marks indicate the calculated Bragg peaks from the experimentally determined crystal structure \cite{EuAg4As2-Sb2}. Despite a small number of unreacted Eu and Sb impurity peaks, there is no evidence of peaks pertaining to EuAgSb. Figures \ref{xray}(b) and (c) are STM images taken at 77 K of the cleaved $ab$ plane, showing a clean surface with a small population of atomic-scale defects. The measured surface lattice constant of $\sim 8.3$ \AA, suggesting a $(\sqrt{3}\times\sqrt{3})R30^\circ$ surface reconstruction.

Figure \ref{property}(a) shows the temperature dependence of the magnetic susceptibility $\chi=M/H$ of \ce{EuAg4Sb2} for $H\parallel c$ (out of plane) and $H\parallel a$ (in plane) under $H=0.1$ T. The inverse of $\chi(T)$ is shown in the inset. Two AFM transitions were observed at $T_2=10.6$ K and $T_1=7.5$ K. The nontrivial low temperature behavior of $\chi(T)$ is highlighted in panel (b). The decrease in $\chi(T)$ in both ordered regions is significantly smaller for $H\parallel c$ comparing to that for $H\parallel a$, suggesting that the hard axis aligns with the $c$ direction. Given its similarity to the envelope seen in \ce{EuAg4As2}, $\chi(T)$ here likely indicates a phase transition from paramagnetic (PM) to an unpinned AFM state at $T_1$, with the material then entering a long-range AFM state at $T_2$. To determine the Curie-Weiss temperature $\Theta_{CW}$ and effective moment $\mu_{eff}$, fits to 1/$\chi(T)$ of both directions using the Curie-Weiss law $1/\chi = \frac{T - \Theta_{CW}}{C}$ are performed, where $C$ is Curie constant and $\mu_{eff}= \sqrt{8C}$. We obtain $\Theta_{CW}^{\parallel c}= 11.6(1)$ K, $\mu_{eff}^{\parallel c}=7.8 \mu_B$/Eu and $\Theta_{CW}^{\parallel a} = 11.0(1) $ K, $\mu_{eff}^{\parallel a}=8.0 \mu_B$/Eu. These values of $\mu_{eff}$ are very close to 7.94 $\mu_B$, the magnetic moment of a free Eu$^{2+}$ cation, indicating that all Eu ions are in the Eu$^{2+}$ state. The positive $\Theta_{CW}$ values are close to that of the polycrystalline value of 12.3 K \cite{HyperfineSplitting}, suggesting strong FM fluctuations in this compound. The anisotropic isothermal magnetization measured at different temperatures is presented in Fig. \ref{property}(c) and (d) for $H\parallel c$ and $H\parallel a$, respectively. The saturation fields are 3.2 T for $H\parallel c$ and 1.1 T for $H\parallel a$, indicating that the $c$ axis is the hard magnetic axis. Multiple metamagnetic transitions were identified and used to map the temperature-field phase diagrams shown in Figs. \ref{property} (e) and (f). For $H\parallel c$, EuAg$_4$Sb$_2$ undergoes sequential transitions through the AFM1 and AFM2 states before reaching the polarized FM state (PFM), while for $H\parallel a$, four distinct AFM states are observed. These complex magnetic phase diagrams highlight the nontrivial AFM states of \ce{EuAg4Sb2}, warranting further investigations, such as neutron scattering, to achieve deeper insights of these magnetic phases presented here. 

\subsection{Quantum Oscillations of \ce{EuAg4Sb2}}

QOs were only observed in the PFM phase, paving the way for a comprehensive understanding of its 3D fermiology. We examined both the temperature dependence and angular dependence of the SdH and dHvA oscillations. For SdH oscillations, $\rho_{yx}$ was measured at various temperatures with the current applied along the $a$ direction. To investigate the dHvA oscillations, we employed magnetic torque, $\vec{\tau} = \vec{M} \times \vec{B}$.

\subsubsection{Temperature Dependence} \label{SdHtemp}

Figure \ref{QOTemp} (a)-(d) summarize SdH oscillations for $H\parallel c$ above 3 T. Figure \ref{QOTemp} (a) shows $\delta\rho_{yx}$, after the subtraction of a polynomial background from $\rho_{yx}$ at various temperatures with the field applied along $c$, as illustrated by the inset of Fig. \ref{QOTemp}(b). The $\rho_{yx}$ curves at temperatures from 2 K - 24 K, before the polynomial background subtraction, can be found in Fig. S1(a) of the supplemental information (SI) \cite{SI}. The Fast Fourier Transform (FFT) seen in Fig. \ref{QOTemp} (b) reveals many features. Four low frequency peaks at $F_{Ha1}$ = 19(5) T, $F_{Hb1}$ = 121(5) T, $F_{Ha2}$ = 155(4) T, and $F_{Hb2}$ = 189(4) T appear, followed by three weaker peaks which are believed to be the 2nd and 3rd higher harmonics of $F_{Ha2}$ and $F_{Hb2}$. Also present at this orientation are two higher frequency peaks $F_{Ta1}$ = 542(9) T and $F_{Tb1}$ = 794(8) T. Figure \ref{QOTemp}(c) shows the temperature-dependent FFT data as a contour map, highlighting the QO peaks with frequencies below 200 T. Upon warming, shifts in frequencies are observed. $F_{Ha1}$ and $F_{Hb1}$ show an increase, while $F_{Ha2}$ and $F_{Hb2}$ exhibit a decrease. 

\begin{figure*}

    \centering
    \includegraphics[width=\textwidth]{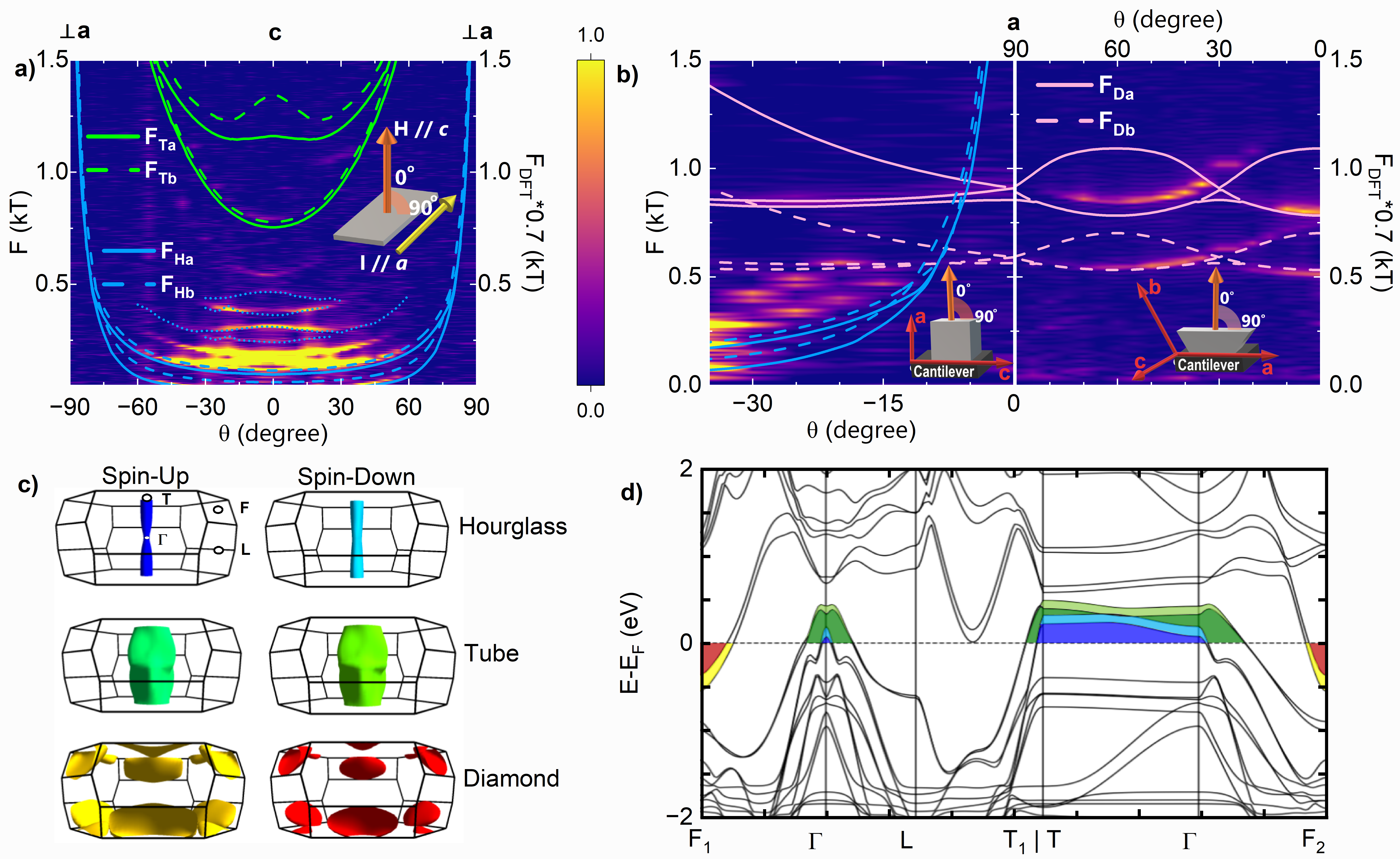}
    \caption{(a) Contour plot of SdH oscillation frequencies and (b) contour plot of the dHvA oscillation frequencies as a function of angle. Solid green, blue, and pink lines represent the DFT computed spin-up frequencies corresponding to the tube, hourglass, and diamond pockets respectively, while dashed lines represent the DFT computed spin-down frequencies. In panel (a), dotted blue lines serve as a guide for the eye to indicate the observed $1^{st}$ and $2^{nd}$ harmonics of $F_{Hb1}$ and $F_{Ha2}$. Insets in each panel illustrate the measurement geometry. (c) 3D renderings of each spin-split component of the Fermi surface are shown separately and annotated with relevant high-symmetry points, for visual clarity. (d) The DFT computed band structure in the PFM state with spins along the $c$ direction.}
    \label{QOContour}
    
\end{figure*}

Figure \ref{QOTemp}(d) summarizes the normalized temperature-dependent FFT amplitudes of $F_{Ha1}$, $F_{Hb1}$, $F_{Ha2}$, $F_{Hb2}$, $F_{Ta1}$, and $F_{Tb1}$. The amplitude in multiple oscillations in resistivity is described by the following adaptation of the Lifshitz-Kosevich (LK) formula:

\begin{equation}
    \frac{\Delta \rho}{{\rho_o}} = \pm \Sigma_i A_i B^{1/2} R_T^i R_D^i R_S^i \: \textrm{cos}[2\pi (\frac{F_i}{B} + \phi_i)]
    \label{LKresist}
\end{equation}

where $A$, $F$, and $\phi$ are constants, frequencies, and phase factors of each pocket respectively \cite{schoenberg}. $R_T = X/ {\rm{sinh}} (X)$ is the thermal damping factor responsible for a finite temperature correction to the Fermi-Dirac distribution, with $X=\alpha T \mu / B {m_e}$, $\alpha =2 \pi^2 K_B m_e/\hbar$ = 14.69 $TK^{-1}$ and $\mu$ = $\frac{m^*}{m_e}$ is the ratio of the cyclotron effective mass to the electron mass. $R_D = {\rm{exp}}(-\alpha T_D \mu / B)$ is the Dingle damping factor, which is related to the quantum lifetime through $\tau_q=\hbar/(2\pi k_BT_D$). $R_S = \rm{cos}  (\pi g \mu/2)$ is the spin damping factor, which accounts for the interference between two oscillations from spin-split Landau levels. 
When multiple frequencies are present, using Eq. \eqref{LKresist} to fit the oscillatory resistivity data and extract the Dingle temperature and the effective mass can be challenging or even impossible. Alternatively, we obtain the effective mass by fitting the temperature-dependent FFT amplitude ($A_{FFT}$) of each peak using $A_{FFT} \propto R_T$, where $B$ is the average inverse field of the FFT window from $B_1$ to $B_2$ and is defined as $1/B = (1/B_1 + 1/B_2)/2$. However, care must be taken when choosing what values to use for $B_1$ and $B_2$, as the wrong choice may lead to an improper estimation of values. For reasons outlined in a previous report on \ce{SrAg4As2}, we used $B_1$ = 5 T and $B_2$ = 9 T \cite{SrAg4As2}. The obtained effective masses are $\mu_{Ha1}$ = 0.10(1), $\mu_{Hb1}$ = 0.06(1), $\mu_{Ha2}$ = 0.11(1), $\mu_{Hb2}$ = 0.077(4), $\mu_{Ta1}$ = 0.26(4), and $\mu_{Tb1}$ = 0.27(5).

 Panels (e) through (h) of Fig. \ref{QOTemp} summarize the dHvA oscillations that were observed in $\tau$ in the PFM phase with applied magnetic fields above 6 T when the magnetic field is applied in the $ab$ plane, 45$\degree$ away from $a$, as shown in the inset of Fig. \ref{QOTemp} (f). Panel (e) illustrates $\delta\tau$ after the subtraction of a polynomial background of $\tau$ at various temperatures to probe the QO. The $\tau$ curves at temperatures from 2 K - 8 K, before the polynomial background subtraction, can be found in Fig. S1(b) of the SI \cite{SI}. The FFT curves outlined in Fig. \ref{QOTemp} (f) reveal two strong frequencies at $F_{Db}$ = 518(6) T and $F_{Da}$ = 813(6) T. No shifts in these frequencies are observed with increasing temperature, as illustrated by the contour map in Fig. \ref{QOTemp} (g), in contrast to the behavior of the small frequency branches shown in Fig. \ref{QOTemp}(c). The amplitude of multiple oscillations in magnetic torque is given by the Lifshitz-Kosevich (LK) theory as

\begin{equation}
    \Delta \tau (B) = \pm \Sigma_i A_i B^{3/2} R_T^i R_D^i R_S^i \: \textrm{sin}[2\pi (\frac{F_i}{B} + \phi_i)]
    \label{osc}
\end{equation}

where the same definitions outlined in the previous section apply \cite{schoenberg}. As before, due to the presence of multiple frequencies, the same procedure implemented for the SdH oscillations was used to extract the effective masses. Figure \ref{QOTemp}(h) shows the normalized temperature-dependent amplitudes to which the thermal damping factor was fit. The extracted effective masses were found to be with $\mu_{Db}$ = 0.31(1) and $\mu_{Da}$ = 0.30(1).

\subsubsection{Angular Dependence}

QO frequency can be shown to be directly proportional to the extreme cross sectional area ($S$) of the Fermi surface perpendicular to the magnetic field through the Onsager relation $F = (\hbar/2\pi e) S$ \cite{schoenberg}. Thus, to obtain the angular dependence of the extreme cross sectional area of each Fermi pocket, we utilized angle-dependent SdH measurements. In this experimental set up, the sample was rotated from the $c$ to $\perp$$a$-axis, as illustrated by the inset of the contour plot of the frequencies in Fig. \ref{QOContour}(a). We observe three branches with similar angle dependence ($F_{Ha1}$ $F_{Hb1}$, and $F_{Hb2}$). Between branches $F_{Hb1}$ and $F_{Hb2}$ there is a branch ($F_{Ha2}$) that does not follow the same angular dependence and rather has a downward curvature that seems to merge with $F_{Hb1}$ at around 30$\degree$. Followed by these branches are a set of three higher harmonics of $F_{Hb1}$ and $F_{Ha2}$ (marked by blue dotted lines). Above 0.5 kT two highly angle dependent branches ($F_{Ta1}$ and $F_{Tb1}$) can be seen. 

\begin{figure*}

    \includegraphics[width=\textwidth]{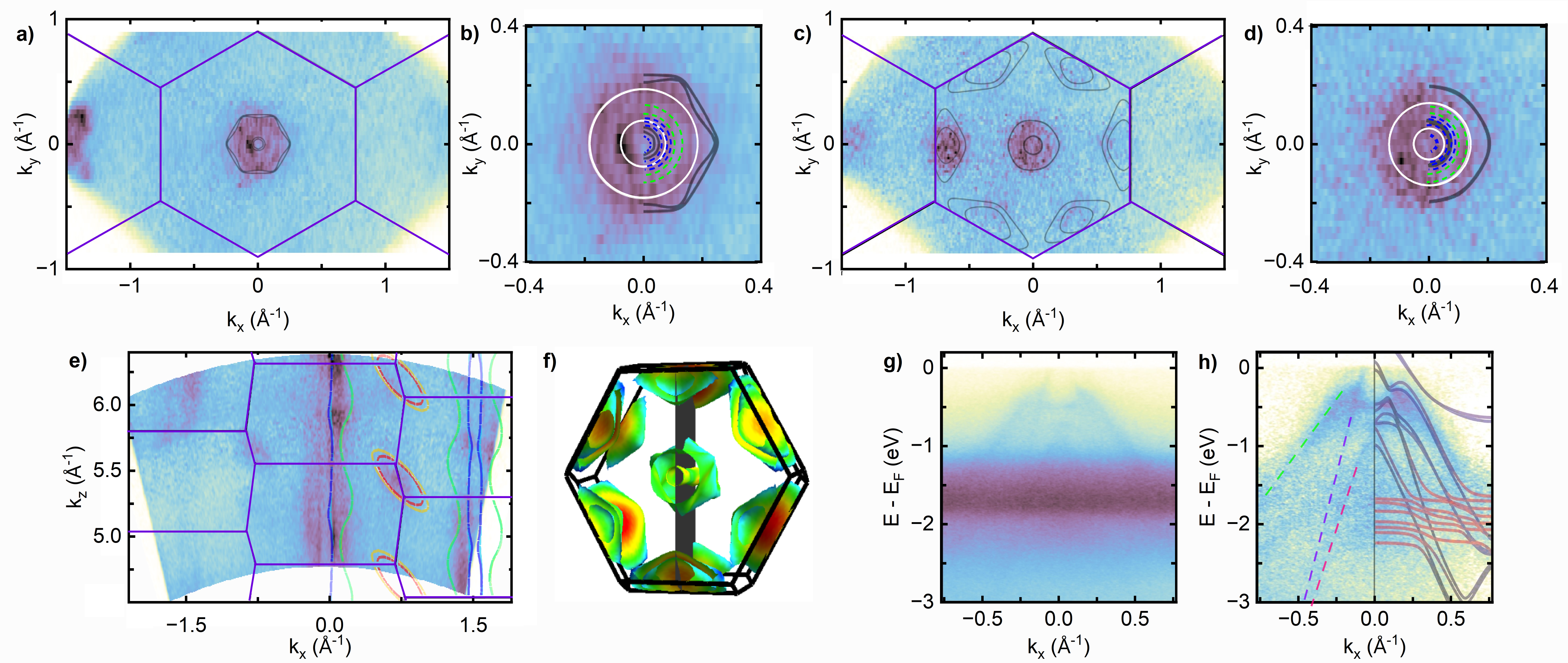}
    \caption{ARPES data taken at 10 K. (a) Fermi surface cut of \ce{EuAg4Sb2}, where data was taken at photon energy $h\nu$ = 100 eV. Solid purple lines indicate the boundaries of the $1^{st}$ Brillouin Zone and solid black lines indicate the DFT-calculated Fermi surface cuts. (b) A zoomed-in view of the feature centered on $k_x=k_y=0$ with data taken at photon energy $h\nu$ = 100 eV. Fit done to the ARPES data: solid white lines; the radii extracted from SdH data: hourglass pocket (dashed blue lines), tube pocket (dashed green lines); DFT calculation: solid black lines. (c) Fermi surface cut at photon energy $h\nu$ = 120 eV. (d) Zoomed-in views of the feature at $k_x=k_y=0$ with data taken at photon energy $h\nu$ = 120 eV. (e) Photon energy map along the M-$\Gamma$-M direction from $h\nu$ = 80 - 160 eV showing $k_z$ dispersion of features at $E_F$. Solid green, blue, and orange/red lines represent DFT computed tube, hourglass, and diamond pockets, respectively. (f) $1^{st}$ Brillouin zone of DFT results showing the (10$\bar{1}$) plane which extends along the F-$\Gamma$-F direction. (g) High-symmetry cut along the M-$\Gamma$-M direction taken at photon energy $h\nu$ = 100 eV. (h) The same high-symmetry cut with non-dispersive high-intensity spectral weight subtracted and DFT computed band structure overlaid. Solid red lines correspond to Eu $f$-orbital projections, while solid black lines are associated with orbitals from Ag and Sb. Green, purple, and pink dashed lines serve as guides for the eye to highlight observed bands. }
    \label{ARPES}
    
\end{figure*}

To complete a 3D view of the Fermi surface, we utilized angle-dependent torque measurements to probe the $ab$ plane. To show continuity in to the $ac$ plane, we also rotated a magnetic field from $a$ to the $c$-axis, as illustrated in the left panel of Fig. \ref{QOContour} (b). Three branches of frequencies can be seen; two lower frequency branches consistent with those in Fig. \ref{QOContour}(a) ($F_{Hb1}$ and $F_{Hb2}$) and a higher frequency branch (F$_{Da}$). The right panel was created by sweeping the magnetic field starting $\perp$ $a$ and moving towards alignment with the $a$-axis. Two fundamental frequency branches, $F_{Da}$ and $F_{Db}$, which exhibit a very similar yet unusual angular dependence, are observed. They clearly demonstrate the presence of spin-split bands arising from the lifting of spin degeneracy caused by magnetism.

\subsection{ARPES}

Figure \ref{ARPES} summarizes the results of our ARPES measurements taken at the MAESTRO beamline at ALS. Figure \ref{ARPES}(a) shows a Fermi surface map taken at 10 K with the photon energy $h\nu = $ \qty{100}{eV}, where data were averaged over $\pm$ \qty{60}{meV} of $E_F$. ARPES intensity at this photon energy is dominated by two roughly circular features centered around $k_x=k_y=0$. Features can also be seen to appear around neighboring Brillouin zones at $k_x=\pm 1.54$ \AA$^{-1}$ = $\pm 2 \pi/a^*$ where $a^*=\sqrt{3}a/2$. A magnification of the circular features centered around $k_x=k_y=0$ is shown in Fig. \ref{ARPES}(b), where data were averaged in a range $\pm$ 100 meV of $E_F$. Through a fit of the raw Momentum Dispersion Curve (MDC) data, which can be found in Fig. S2 of the SI \cite{SI}, the features were found to have radii of 0.08(3) \AA$^{-1}$ and 0.19(2) \AA$^{-1}$ and are indicated by overlaid white solid lines. Data were also taken at $h\nu$ = 120 eV and are shown in Fig. \ref{ARPES}(c). In addition to two circular features centered at $k_x=k_y=0$, ellipsoidal features were seen surrounding the circular features in a ring-like formation. Figure \ref{ARPES}(d) illustrates a magnified view of the circular features centered around $k_x=k_y=0$. Fitting the raw MDC data, reveals radii of 0.05(4) \AA$^{-1}$ and 0.14(7) \AA$^{-1}$, again indicated by overlaid white solid lines. 

To further investigate the $k_z$ dispersion of the features seen in the two Fermi surface maps, we performed a photon energy mapping along the M-$\Gamma$-M high-symmetry direction over the range $h\nu$ = 80 - 160 eV. The data were converted to momentum-space assuming an inner potential of $V_0=6$ eV inferred from the spacing of periodic features in the $k_z$ map and can be seen in Fig. \ref{ARPES}(e). Two nested features, continuous along $k_z$ and centered at $k_x=0$, can be seen as well as two similar features centered around neighboring Brillouin zones at $k_x = \pm 1.54$ \AA$^{-1}$ = $\pm 2 \pi/a^*$. The inner feature is nearly non-dispersive with $k_z$, while the outer feature exhibits a slight periodic modulation along $k_z$. Small ellipsoidal features also appear between the repeating cylindrical features along $k_x$.

To probe the $k_x$ dispersion of the bands, a high-symmetry cut of the ARPES band structure at photon energy $h\nu$=100 eV along the M-$\Gamma$-M direction is shown in Fig. \ref{ARPES}(g). The spectral weight is dominated by a region of high-intensity near 2 eV binding energy that is virtually non-dispersive with in-plane momentum $k_\parallel$. To minimize the effect of these non-dispersive bands and better investigate the dispersive features, we perform a normalization to remove them by subtracting the median Energy Dispersion Curve (EDC) over the entire region in a method adapted from work done on MnTe \cite{Krempasky}. The resulting ARPES \textit{E-k} image can be seen in Fig. \ref{ARPES}(h) with three pronounced features. Two inner bands with approximately linear dispersion can be seen, where one reaches an apex at approximately 1 eV (pink dashed line) and the other appears to cross $E_F$ to form a hole-like Fermi pocket (purple dashed line). The third outer band (green dashed line) also appears near $E_F$ and seems to form a larger hole-like pocket.

\begin{table*}[ht]
\caption{Parameters extracted from dHvA data at two different field orientations. $\alpha$: $45\degree$ from $a$, as shown in the inset of Fig. \ref{QOTemp}(f). Errors associated with frequencies F were found via the Full Width Half Maximum technique. While errors for $m^*_{exp}$ were found by computing the variance-covariance matrix from the fitting function produced from fitting the data with the Levenberg-Marquardt (L-M) Algorithm in Origin. All other errors were found by way of error propagation.}
\label{meff}
\centering
\begin{tabular}{ c| c| c | c | c | c | c | c }
 \hline
 \hline
           &$B$ & $F$(T) & $m^*_{\rm{exp}} (m_e)$  & $m^*_{\rm{DFT}} (m_e)$  & $k_F$(\AA$^{-1}$) & $v_F$ ($10^{5}$m/s) & ${p_{\rm{QO}}(10^{26}/\si{m^{3}})}$  \\
 \hline
 {$F_{Ha1}$}  & \multirow{4}{*}{$\parallel c$}& 19(5)  & 0.10(1) & 0.082  & 0.02(1) & 2(1) & \multirow{3}{*}{4.1(4)}  \\
 {$F_{Ha2}$}  &  & 155(4)  & 0.11(1) & 0.091  & 0.07(1) & 7(1) &  \\
 {$F_{Hb1}$}  &  & 121(4)  & 0.06(1) & 0.080  & 0.06(1) & 12(3) & \\
 {$F_{Hb2}$}  &  & 189(4)  & 0.077(4) & 0.091  & 0.08(1) & 12(2) &\\
 \hline
 {$F_{Ta1}$}  & \multirow{2}{*}{$\parallel c$ } & 542(9)  & 0.26(4) & 0.38 & 0.13(2)  & 6(1) & \multirow{2}{*}{0.71(8)}  \\      
 {$F_{Tb1}$} &  & 794(8) & 0.27(5) & 0.36 & 0.16(2)  & 7(2) &  \\
 \hline   
 {$F_{Da}$} & \multirow{2}{*}{$\parallel \alpha$} & 813(6) & 0.30(1) & 0.31 & 0.16(1) & 6.1(4) &   \\
 {$F_{Db}$} &  & 518(6)  & 0.31(1) & 0.28 & 0.13(1) & 4.8(4)  &  \\
\hline
\hline

\end{tabular}
\label{table1}
\end{table*}

\section{Discussion} 

\subsection{Comparison to DFT}

To better understand the QO angular dependence, DFT calculations were performed. First, the structure of \ce{EuAg4Sb2} was optimized in the the primitive cell and found to be in good agreement with previous results (see table S1)\cite{SI, EuAg4As2-Sb2}. Our calculations were performed in the PFM state with magnetic moments oriented along the $c$ direction, as all QOs were observed in the PFM state, as shown in Figs. 2 and 3. On-site Coulomb $U$ parameters of $4, 6,$ and 8 eV were used. Figure \ref{QOContour}(c) and (d) show the Fermi surface and band structure obtained when using $U=6$ eV while Fig. S3 presents the band structures calculated with $U = 4$ and $U = 8$ eV \cite{SI}. No significant changes are observed in the size and shape of the bands at the Fermi level when varying $U$. However, as $U$ is increased the Eu flat bands are pushed further below the fermi level and when decreased they migrate closer to the fermi level. Our ARPES data  (Fig. \ref{ARPES}(g)) shows a clear signal of the Eu flat bands near -2 eV, which is consistent with our choice of U = 6 eV. Therefore, we conclude that $U = 6$ eV is a reasonable choice. Notably, $U=6$ eV was also chosen for other Eu containing materials \cite{DFT+U1, DFT+U2}.

The density of states at the Fermi level are dominated by Ag and Sb orbitals. The Fermi Surface contains three pockets that are spin-split: An hourglass-shaped, long, continuous hole pocket oriented along $c$; a tubular one, which is a wider, long and continuous hole pocket also oriented along $c$ that encloses the hourglass pocket; and a set of three diamond-shaped electron pockets centered at the F points on the faces of the 1st Brillouin zone.

Quantum oscillation frequencies associated with each Fermi pockets were calculated and compared to SdH and dHvA measurement. Solid lines overlaid on the contour plots in Fig. \ref{QOContour}(a) and (b) are the DFT computed spin-up frequencies corresponding to the different pockets and dashed lines represent the spin-down frequencies. A scaling factor of 0.7 is applied in order to maximize quantitative similarity. The minimum of $F_H$ and $F_T$ in the $c$-$ab$ sweep (Fig. \ref{QOContour}(a)) occur when the field lies parallel to $c$, $\textit{i.e.}$ when the field is applied along the length of the hourglass and tube pockets (Fig. \ref{QOContour}(c)). A maximum frequency value is reached when the field is rotated 87$\degree$ and 80$\degree$ away from the $c$-axis for the hourglass and tube pockets respectively. Due to the continuous nature of the pockets along $c$ and thus an infinite extreme cross section perpendicular to the $ab$ plane, the frequencies are expected to diverge before the field reaches the $ab$ plane and not appear when the field is swept in this plane. This expectation is consistent with what was observed in the right panel of Fig. \ref{QOContour}(b). The multiplicity and lower symmetry of the diamond pocket create the more complex $F_D$ profiles in both the $a$-$c$ and $ab$ sweeps presented in Fig. \ref{QOContour}(b).

Overall, good agreement was achieved between the experimental and DFT-computed angle dependencies of QO frequencies despite some fine features for the very small hourglass pocket. Small quantitative deviations can occur due to the approximate nature of DFT \cite{skeaf}. Additionally, although crystalline defects in \ce{EuAg4Sb2} are expected to be low \cite{HyperfineSplitting}, this could contribute to a shift of the Fermi level, which can lead to sizable changes for tiny pockets. With these factors in mind, we conclude that the essential features of the Fermi surface are correctly described by our DFT calculations.  

 Based on our QO data and the shape of the Fermi surface determined from DFT, we can estimate the hole carrier densities associated with the tube and hourglass pockets. Assuming circular cross sections, the Fermi wave vector can be calculated from the Onsager relation as $k_F = \sqrt{2eF/\hbar}$. Wave vectors associated with the two tube pockets are $k_{Ta1} =0.13(2)$ \AA$^{-1}$ and $k_{Tb1}=0.16(2)$ \AA$^{-1}$ with Fermi velocities $v_F = \hbar k_F/m^*$ valued at $v_{{Ta1}}=6(1) \times 10^5~\si{m/s}$ and $v_{{Tb1}}=7(2) \times 10^5~\si{m/s}$. Values used for $F_{Ta1}$, $F_{Tb1}$, and their corresponding effective masses were taken from data collected when the field is parallel to the $c$ axis (Table \ref{table1}). The carrier density of a tube pocket with only spin-up or spin-down electrons can be estimated by $\pi k_T^2 (2\pi/c^*)/(8\pi^3)$ where $c^*=c/3$, resulting in the total carrier density of the two tube pockets as $p_{{T}}= 4.1(4) \times 10^{26}$ m$^{-3}$. We can calculate the carrier density associated with the hourglass pocket by following the same method. Values described in Table \ref{table1} for $F_{Ha1}$, $F_{Hb1}$, $F_{Ha2}$, $F_{Hb2}$, and their associated effective masses were used. The corresponding wave vectors were found to be $k_{Ha1}=0.02(1)$ \AA$^{-1}$, $k_{Ha2}=0.07(1)$ \AA$^{-1}$, $k_{Hb1}=0.06(1)$ \AA$^{-1}$, and $k_{Hb2}=0.08(1)$ \AA$^{-1}$. The Fermi velocity are then $v_{Ha1} = 2(1) \times 10^5~\si{m /s}$, $v_{Ha2} = 7(1) \times 10^5~\si{m /s}$, $v_{Hb1} = 12(3) \times 10^5~\si{m /s}$ and $v_{Hb2} = 12(2) \times 10^5~\si{m /s}$. By approximating the hourglass pockets to be a pair of cylinders with radius $k_{Ha}=(k_{Ha1}+k_{Ha2})/2$ and $k_{Hb}=(k_{Hb1}+k_{Hb2})/2$ we find the carrier density of them to be $p_H =0.71(8) \times 10^{26} $ m$^{-1}$ and thus the total hole carrier density obtained from QO is estimated as $p_{\rm{QO}} =4.8(4) \times 10^{26} $ m$^{-3}$. These findings classify \ce{EuAg4Sb2} as an AFM semimetal. 

\subsection{Comparison to ARPES}
To further verify the DFT results, the calculations were compared to ARPES data. The main results are shown in Fig. \ref{ARPES}. In panels (a)-(d), solid black lines represent the DFT-computed cross sections of the Fermi pockets, and solid purple lines indicate the Brillouin zone boundaries. Data taken at a photon energy of 100 eV (Fig. \ref{ARPES}(a) and (b)) was found to correspond to $k_z$ = 5.27 \AA$^{-1}$ $\sim$ 6.9 $\times$ $(2\pi/c^*)$ where $c^*=c/3$, \textit{i.e.} the cross section near the $\Gamma$ point, where no diamond pockets are expected to exist. Indeed, no features at the edges of the Brillouin zone are seen. Data taken at a photon energy of 120 eV (Fig. \ref{ARPES}(c) and (d)) was found to correspond to $k_z$ = 5.75  \AA$^{-1}$ $\sim$ 7.5 $\times$ $(2\pi/c^*)$ \textit{i.e.} the cross section near the T point. At this $k_z$ cut, it is expected that the diamond pockets will be visible at the edge of the Brillouin zone and the tube and hourglass pockets at the center. Indeed, significant spectral weight resembling the expected shapes of the pockets can be seen at the center and appropriate locations at the edges. In panels (b) and (d), radii extracted from SdH oscillation measurements associated with the hourglass and tube pockets are overlaid and represented by blue and green dashed semi circles, respectively. Good agreement is seen between the predicted sizes of the hourglass and tube features identified by DFT and the radii observed in the ARPES spectra. Next, various vertical planes of the computed Fermi surface in the first Brillouin zone were projected in 2D and compared to the ARPES photon energy dispersion map as shown in Fig. \ref{ARPES}(e). Upon overlaying the Fermi surface cut along the (10$\bar{1}$) plane, significant spectral weight is observed within the bounds of the overlaid pockets, represented by solid green, blue, and orange/red lines for the tube, hourglass, and diamond pockets, respectively. This alignment illustrates good agreement between the DFT predictions and the ARPES data. Finally, a comparison of the DFT band structure and the ARPES measurement was made by overlaying the band structure along $\Gamma$-L onto the high symmetry cut along M-$\Gamma$-M direction of the ARPES data taken at h$\nu$=100 eV (Fig. \ref{ARPES}(g) and (h)). This comparison reveals good agreement with the approximately linearly dispersing features observed in ARPES data. The highly non-dispersive feature with large spectral weight seen near -2 eV was found to match well with a collection of flat bands arising from Eu $f$-orbitals, indicated by the red bands in Fig. \ref{ARPES}(h). Through this comparison, we further conclude that the Fermi surface and band structure of \ce{EuAg4Sb2} are correctly described by our DFT calculations.

\subsection{The effect of magnetism on the electronic structure}
As shown in Fig. \ref{QOTemp}(c), the small hourglass pocket experiences significant changes in size as the temperature increases, gradually transforming into a more cylindrical shape. This is in sharp contrast with its nonmagnetic analog \ce{SrAg4Sb2} which exhibits no temperature-dependent shifts in QO peaks \cite{SrAg4As2}, as illustrated in Fig. S4 \cite{SI}. The distinct behavior suggests that the shifts are not driven by temperature alone but are instead closely linked to magnetism. However, there is no indication that the shifts are directly caused by the different magnetic phases below 7.5 K or between 7.5 K and 10.6 K. For instance, as shown in Fig. \ref{QOTemp}(c), although $F_{Hb1}$ and $F_{Ha1}$ became too weak to detect above 10 K, they exhibit a monotonic increase without notable features at 7.5 K. Similarly, both $F_{Hb2}$ and $F_{Ha2}$ monotonically decrease from 2 K to 24 K, showing no distinct features at 7.5 K or 10.6 K. This is not surprising since all QOs emerge in the PFM phase from 2 K to 24 K. The aspect of magnetism in the PFM phase that changes upon warming is the saturation moment, which reduces due to thermal fluctuations. Therefore, the observed QO shifts highlight the sensitivity of the small Fermi pockets to variations in exchange splitting caused by changes in magnetic moment. Modifications in exchange interaction between electrons and 4$f$ local moments have been seen to cause magnetism-induced temperature-dependent shifts in QO peaks in several magnetic Weyl, Dirac semimetals with small Fermi pockets, including \ce{CeBiPt}, \ce{NdAlSi}, \ce{EuMnSb2}, \ce{EuMnBi2}, and \ce{PrAlSi} \cite{CeBiPt, NdAlSi-2023, NdAlSi-2021, EuMnSb2-2021, EuMnSb2-2023, EuMnBi2, PrAlSi}. In all of these compounds, Fermi pockets with frequencies below 200 T were noticeably affected. We further notice that the larger Fermi pockets, for example, the diamond ones, remain unaffected, as shown in Fig. 3(g). This suggests that while variations of the exchange splitting can impact the tiny Fermi pockets, they have a negligible effect on the relatively larger ones. To investigate this further, DFT calculations were performed in the PFM state with the moments oriented along a different direction (Fig. S5 in SI), as well as in a simple AFM state (Fig. S6 in SI) \cite{SI}. Indeed, across all these magnetic models, the overall band structure near the Fermi level and the resulting Fermi surfaces remain largely unchanged, aside from some subtle variations in the features of the hourglass pocket. The robustness of the band structure of \ce{EuAg4Sb2} near the Fermi level with respect to the choice of $U$ and magnetic configurations is likely due to the dominance of the weakly correlated $s-$ and $p-$ orbits of Ag and Sb atoms in the bands around the Fermi level, while the strongly-correlated Eu flat bands are -2 eV below.

\section{Conclusion}

In summary, we have investigated the 3D fermiology of \ce{EuAg4Sb2} using quantum oscillation, ARPES and DFT calculations. Temperature-dependent SdH and dHvA oscillations reveal small effective masses associated with the Fermi pockets. Angle-dependent SdH and dHvA oscillations along with ARPES data show good agreement with DFT calculations performed at the PFM state, revealing spin-split hourglass-shaped hole pockets and spin-split tube-shaped hole pockets, both centered at the $\Gamma$ point, and spin-split diamond-shaped electron pockets centered at the F points. Furthermore, no significant changes in band structure calculations near the Fermi level were observed across FM or simple AFM configurations.

\section*{Acknowledgments}
N. N. thanks Prof. Igor I. Mazin for the fruitful discussion. Work at UCLA was supported by the U.S. Department of Energy (DOE), Office of Science, Office of Basic Energy Sciences under Award Number DE-SC0021117. M. M and H. C. acknowledge the support from U.S. DOE BES Early Career Award KC0402010 under Contract No. DE-AC05-00OR22725 and the U.S. DOE, Office of Science User Facility operated by the ORNL. This research used resources of the Advanced Light Source, which is a DOE Office of Science User Facility under contract no. DE-AC02-05CH11231. The computational work was supported by the DOE-BES grant DE-SC0024987 to A. N. A., and used the Hoffman2 Shared Cluster provided by UCLA Office of Advanced Research Computing's Research Technology Group.

\medskip

\bibliographystyle{apsrev4-1}
\bibliography{Eu142}

\end{document}